\documentclass[a4paper,12pt]{elsart}
\usepackage{epsfig}
\usepackage{array,cite}
\newcommand{\newc}{\newcommand}
\newc{\definmath}[2] {\def#1{\ifmmode#2\else$#2$\fi}}

\definmath\gsim{\,\,\rlap{\raise 3pt\hbox{$>$}}{\lower 3pt\hbox{$\sim$}}\,\,}
\definmath\lsim{\,\,\rlap{\raise 3pt\hbox{$<$}}{\lower 3pt\hbox{$\sim$}}\,\,}

\definmath\amin{\mathrm{min}}
\definmath\amax{\mathrm{max}}

\def\ie{{\it{i.e.}}}

\def\compProg{\tt}
\def\herwig{{\compProg HERWIG}}
\def\herwigv#1{{\compProg HERWIG-#1}}
\def\isajet{{\compProg ISAJET}}
\def\isajetv#1{{\compProg ISAJET-#1}}

\def\atlfastv#1{{\compProg ATLFAST-#1}}

\newc{\barr}{\begin{eqnarray}}
\newc{\earr}{\end{eqnarray}}
\newc{\beq}{\begin{equation}}
\newc{\eeq}{\end{equation}}

\newc{\voidcol}{\phantom{m}\begin{rotate}{270}Not
reconstructed\end{rotate}\phantom{mn}}

\definmath\half{{\frac 1 2}}
\definmath\threehalfs{{\frac 3 2}}
\definmath\quarter{{\frac 1 4}}
\definmath\sixth{{\frac 1 6}}
\definmath\third{{\frac 1 3}}
\definmath\twothirds{{\frac 2 3}}
\definmath\fourthirds{{\frac 4 3}}
\definmath{\mPl}{M_\mathrm{Pl}}
\definmath{\invfb}{\mathrm{fb}^{-1}}
\definmath{\nsec}{\mathrm{ns}}

\definmath{\Omegadm}{\Omega_\mathrm{CDM}}
\definmath{\omegadm}{\omega_\mathrm{CDM}}

\definmath{\micron}{{\mu\mathrm{m}}}

\definmath{\silica}{{\mathrm{SiO_2}}}

\definmath{\mnought}{{m_0}}
\definmath{\mhalf}{{m_\frac{1}{2}}}
\definmath{\mthreehalfs}{{m_{3/2}}}
\definmath{\DeltaMChi}{{\Delta M_{\cht_1}}}
\definmath\mtx{{m_{TX}}}
\definmath\mttwo{{m_{T2}}}
\definmath\mttwosq{{m_{T2}^2}}
\definmath\to{\rightarrow}
\newc{\mr}{\mathrm}

\def\GeV{{{\mr {GeV}}}}
\def\TeV{{{\mr {TeV}}}}

\def\slashchar#1{\setbox0=\hbox{$#1$}           % set a box for #1
   \dimen0=\wd0                                 % and get its size
   \setbox1=\hbox{/} \dimen1=\wd1               % get size of /
   \ifdim\dimen0>\dimen1                        % #1 is bigger
      \rlap{\hbox to \dimen0{\hfil/\hfil}}#1 
   \else                                        % / is bigger
      \rlap{\hbox to \dimen1{\hfil$#1$\hfil}}/                                    \fi}

\definmath{\etmiss}{\slashchar{E}_T}
\definmath{\pmiss}{\slashchar{p}}
\definmath{\ptmiss}{\slashchar{p}_T}
\definmath{\Ptmiss}{\slashchar{{\bf p}}_T}
\definmath{\pt}{p_T}
\definmath{\qth}{Q_\mr{thr}}

\newc{\appref}[1]{appendix~\ref{#1}}
\newc{\chref}[1]{chapter~\ref{#1}}
\newc{\Chref}[1]{Chapter~\ref{#1}}
\newc{\secref}[1]{section~\ref{#1}}
\newc{\eqref}[1]{eq.~\ref{#1}}
\newc{\tabref}[1]{table~\ref{#1}}
\newc{\figref}[1]{fig.~\ref{#1}}
\newc{\figsref}[1]{figs.~\ref{#1}}
\newc{\partref}[1]{part~\ref{#1}}
\newc{\Secref}[1]{Section~\ref{#1}}
\newc{\Eqref}[1]{Eq.~\ref{#1}}
\newc{\Tabref}[1]{Table~\ref{#1}}
\newc{\Figref}[1]{Fig.~\ref{#1}}
\newc{\Partref}[1]{Part~\ref{#1}}

\definmath{\z}  {\mathrm{Z}^{0}}
\definmath\tbar{{\bar t}}
\definmath\ttbar{{t \tbar}}

\definmath{\cht}{\tilde{\chi}}
\definmath{\chgone}{{\cht^+_1}}
\definmath{\chgtwo}{{\cht^+_2}}
\definmath{\chgonem}{{\cht^-_1}}
\definmath{\chgonepm}{\cht^{\pm}_1}
\definmath{\chgall}{\cht^{\pm}_{1,2}}
\definmath{\ntlone}{{{\cht^0_1}}}
\definmath{\ntltwo}{{{\cht^0_2}}}
\definmath{\ntlthree}{\cht^0_3}
\definmath{\ntlfour}{\cht^0_4}
\definmath{\ntlall} {\tilde{\chi}_{1,2,3,4}^{0}}
\definmath{\gluino}{\tilde{g}}

\definmath{\ssul} {{\tilde{u}_{L}}}
\definmath{\ssdl} {{\tilde{d}_{L}}}
\definmath{\sscl} {\tilde{c}_{L}}
\definmath{\sssl} {\tilde{s}_{L}}
\definmath{\sstone} {\tilde{t}_{1}}
\definmath{\ssbone} {\tilde{b}_{1}}
\definmath{\ssur} {\tilde{u}_{R}}
\definmath{\ssdr} {\tilde{d}_{R}}
\definmath{\sscr} {\tilde{c}_{R}}
\definmath{\sssr} {\tilde{s}_{R}}
\definmath{\ssttwo} {\tilde{t}_{2}}
\definmath{\ssbtwo} {\tilde{b}_{2}}
\definmath{\squark} {{\tilde{q}}}
\definmath{\sqr} {\tilde{q}_{R}}
\definmath{\sql} {{\tilde{q}_{L}}}
\definmath{\squark} {{\tilde{q}}}
\definmath{\ssulbr} {\bar{\tilde{u}_{L}}}
\definmath{\ssdlbr} {\bar{\tilde{d}_{L}}}
\definmath{\ssclbr} {\bar{\tilde{c}_{L}}}
\definmath{\ssslbr} {\bar{\tilde{s}_{L}}}
\definmath{\sstonebr} {\bar{\tilde{t}_{1}}}
\definmath{\ssbonebr} {\bar{\tilde{b}_{1}}}

\definmath{\ssurbr} {\bar{\tilde{u}_{R}}}
\definmath{\ssdrbr} {\bar{\tilde{d}_{R}}}
\definmath{\sscrbr} {\bar{\tilde{c}_{R}}}
\definmath{\sssrbr} {\bar{\tilde{s}_{R}}}
\definmath{\ssttwobr} {\bar{\tilde{t}_{2}}}
\definmath{\ssbtwobr} {\bar{\tilde{b}_{2}}}

\definmath{\ssel} {\tilde{e}_{L}}
\definmath{\ssellp} {\tilde{e}_{L}^{+}}
\definmath{\ssellm} {\tilde{e}_{L}^{-}}
\definmath{\ssellpm} {\tilde{e}_{L}^{\pm}}

\definmath{\sser} {\tilde{e}_{R}}
\definmath{\sselrp} {\tilde{e}_{R}^{+}}
\definmath{\sselrm} {\tilde{e}_{R}^{-}}
\definmath{\sselrpm} {\tilde{e}_{R}^{\pm}}

\definmath{\ssmulp} {\tilde{\mu}_{L}^{+}}
\definmath{\ssmulm} {\tilde{\mu}_{L}^{-}}
\definmath{\ssmulpm} {\tilde{\mu}_{L}^{\pm}}

\definmath{\ssmurp} {\tilde{\mu}_{R}^{+}}
\definmath{\ssmurm} {\tilde{\mu}_{R}^{-}}
\definmath{\ssmurpm} {\tilde{\mu}_{R}^{\pm}}

\definmath{\sstauone} {{\tilde{\tau}_{1}}}
\definmath{\sstauonep} {\tilde{\tau}_{1}^{+}}
\definmath{\sstauonem} {\tilde{\tau}_{1}^{-}}
\definmath{\sstauonepm} {\tilde{\tau}_{1}^{\pm}}

\definmath{\sstautwop} {\tilde{\tau}_{2}^{+}}
\definmath{\sstautwom} {\tilde{\tau}_{2}^{-}}
\definmath{\sstautwopm} {\tilde{\tau}_{2}^{\pm}}

\definmath{\sslrpm} {{\tilde{l}_{R}^{\pm}}}
\definmath{\sslr} {{\tilde{l}_{R}}}
\definmath{\ssll} {{\tilde{l}_{L}}}

\definmath{\ssnu} {\tilde{\nu}}
\definmath{\ssnuel} {\tilde{\nu}_{e}}
\definmath{\ssnumul} {\tilde{\nu}_{\mu}}
\definmath{\ssnutl} {\tilde{\nu}_{\tau}}

\definmath{\lqnear} {{l^\mathrm{near}q}}
\definmath{\lqfar} {{l^\mathrm{far}q}}
\definmath{\lqhigh} {{l^\mathrm{high}q}}
\definmath{\lqlow} {l{^\mathrm{low}q}}

\definmath{\lqbnear} {{l^\mathrm{near}\bar{q}}}
\definmath{\lqbfar} {{l^\mathrm{far}\bar{q}}}
\definmath{\lqbhigh} {{l^\mathrm{high}\bar{q}}}
\definmath{\lqblow} {{l^\mathrm{low}\bar{q}}}

\definmath{\mlqnear}{{m_{lq}^\mathrm{near}}}
\definmath{\mlqnearsq}{{\left(m_{lq}^\mathrm{near}\right)^2}}
\definmath{\mlqnearmax}{{\left(\mlqnear\right)_\mathrm{max}}}
\definmath{\mlqnearmaxsq}{{\left(\mlqnear\right)_\mathrm{max}^2}}

\definmath{\mlqbnear}{{m_{l\bar{q}}^\mathrm{near}}}
\definmath{\mlqfar}{{m_{lq}^\mathrm{far}}}
\definmath{\mlqbfar}{{m_{l\bar{q}}^\mathrm{far}}}

\definmath{\lqplus} {{l^+q}}
\definmath{\lqminus} {{l^-q}}

\definmath{\lamp}{\lambda^\prime}
\definmath{\lampp}{\lambda^{\prime \prime}}
\definmath{\rparity}{{R_P}}

\newif\iftth
\iftth\else
\newc{\prepareAbbrev}[7]{\newcounter{#5}\newcommand{#1}{\mygloss{#2}{#6}{#4 #7}\ifnum\arabic{#5}=0 {#4 (#3)}\else#3\fi\addtocounter{#5}{1}}}
\fi

\def\anti{\mbox{(anti)}\-}

\newcommand{\EPSFIGURE}[3][v]{\begin{figure}[#1]\begin{center}\epsfig{file=#2}
                                        \caption{#3}\end{center}\end{figure}}
\def\letter{{Letter}}
\begin{document}

\begin{frontmatter}

\title{
Determining the spin of supersymmetric particles at the {LHC}
using lepton charge asymmetry.
}

\author{A.J. Barr}
\address{Cavendish Laboratory, University of Cambridge, Madingley Road,
        Cambridge, CB3 0HE, UK}

\ead{alan.barr@cern.ch}
\begin{abstract}
If signals suggesting supersymmetry (SUSY) are discovered at the LHC then
it will be vital to measure the spins of the new particles to 
demonstrate that they are indeed the predicted super-partners.
A method is discussed by which the spins of some of the SUSY
particles can be determined.
Angular distributions in sparticle decays
lead to charge asymmetry in 
lepton-jet invariant mass distributions.
The size of the asymmetry is proportional to the 
primary production asymmetry between squarks and anti-squarks.
Monte Carlo simulations are performed for a particular 
mSUGRA model point at the LHC.
The resultant asymmetry distributions are consistent with 
a \mbox{spin-0} slepton and a spin-\half\ \ntltwo, 
but are not consistent with both particles being scalars.

\end{abstract}
\begin{keyword}
Hadronic Colliders, Supersymmetry, Spin, LHC
\\Cavendish HEP-2004-14
\end{keyword}

\end{frontmatter}

\section{Spin correlations and charge asymmetry}

\EPSFIGURE{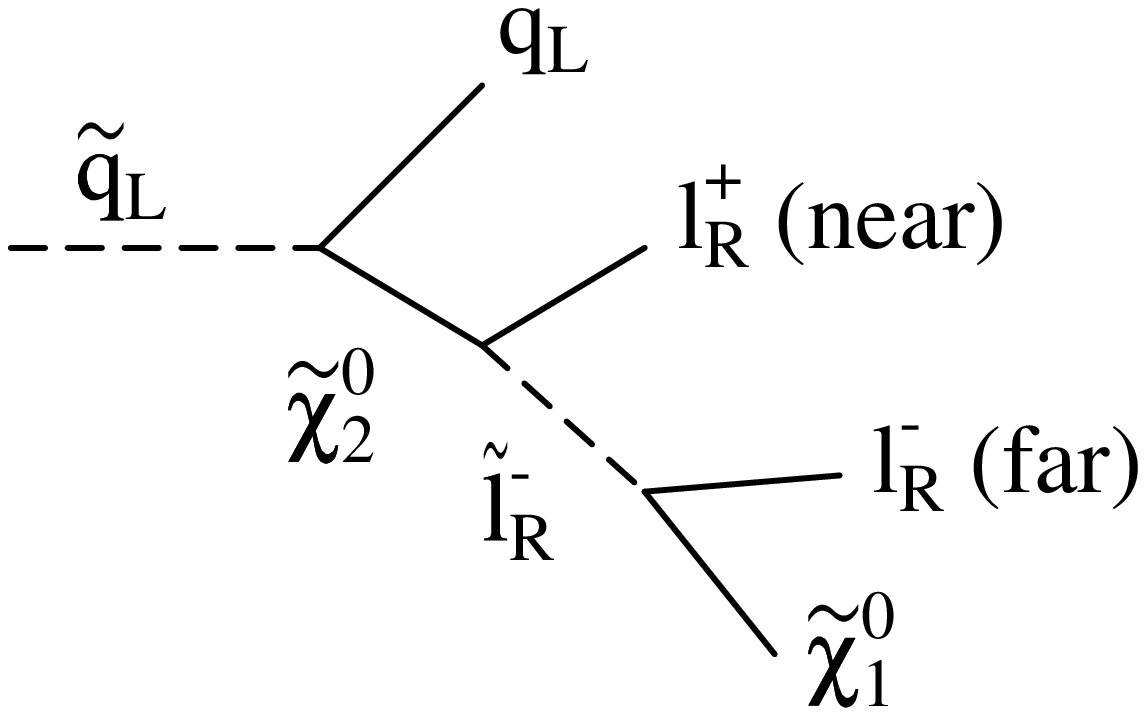, height=4cm}{
The decay chain under investigation. The lepton from the \ntltwo\ decay 
is labeled the `near' lepton, regardless of its charge.
Also considered are the diagrams after charge conjugation of the 
slepton and leptons; or of the squark and quark; 
or of the entire diagram.
\label{fig:feyn}
}

A recent publication\cite{Richardson:2001df} describes the method by which 
spin correlations were added to the \herwig\cite{Corcella:2000bw,Moretti:2002eu}
Monte Carlo event generator. It includes an example
of part of a supersymmetric decay chain, 
\begin{equation}
\label{eq:partchain}
\sql \to \ntltwo q_L \to \sslrpm l^\mp q_L
\end{equation}
in which spin correlations can play a significant role in
the kinematics of the emitted particles.
When the decay of the slepton is also considered, 
(\figref{fig:feyn}), the final state consists of two opposite-signed
leptons of the same family, a quark jet, and missing energy from the
undetected \ntlone.
%The Feynman diagram in which the anti-slepton participates
%in the decay chain is shown in \figref{fig:feyn}~b.

\EPSFIGURE{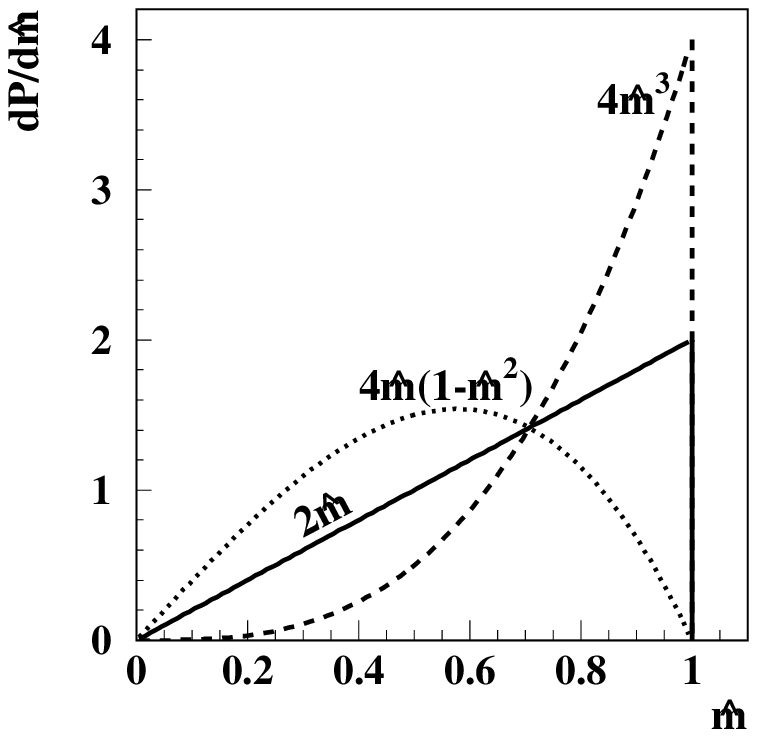, height=7cm}{
Idealised shapes of the \lqnear\ invariant mass distributions
in terms of the re-scaled invariant mass variable
$\hat{m}$ defined in eq.~\ref{eq:mhat}.
The solid line shows $\frac{dP_\mathrm{PS}}{d\hat{m}}$ (\eqref{eq:Pps}),
the dashed line $\frac{dP_1}{d\hat{m}}$ (\eqref{eq:P1}),
and the dotted line $\frac{dP_2}{d\hat{m}}$ (\eqref{eq:P2}).
\label{fig:shapes}
}

%Of course this information is not all available to the experimental 
%physicist, but it is simpler to understand how the 
%spin correlations effect these distributions.
In this \letter\ I examine the observability of spin effects in this decay chain at the LHC.
I first consider the distribution of invariant mass of the quark 
or anti-quark jet from the squark decay, 
and the lepton or anti-lepton from the \ntltwo\ decay.
Using the terminology of \cite{Lester:2001zx}, 
this lepton is referred to as the `near' lepton, as opposed to the lepton
from the slepton decay, which is called the `far' lepton.
Initially I investigate the parton-level distributions which 
one would obtain if the sign of the squark and the identity 
of the lepton were known.
In \secref{sec:plusminus} the experimentally observable 
distributions are introduced. 
These have added complications due to the difficulty in 
(a) distinguishing which lepton came from which decay, and 
(b) measuring the charge of the (s)quark.

In the approximation in which the quark and lepton are massless,
and the SUSY particles are on-mass-shell, then the \lqnear\ invariant mass
has a simple and direct interpretation in terms of
the angle, $\theta^*$, between the quark and lepton in the $\ntltwo$
rest frame,
\begin{equation}
\mlqnearsq %%= ( p_l + p_q )^2 = p_l^2 + p_q ^2 + 2 (p_l\cdot p_q)
            = 2 |{\bf p}_l| |{\bf p}_q|(1-\cos\theta^*)
            = \mlqnearmaxsq \sin^2(\theta^*/2)
\end{equation}
where ${\bf p}_l$ and ${\bf p}_q$ are the 3-momenta of the near lepton and the
quark respectively in the \ntltwo\ rest frame, and the kinematic maximum of \mlqnear\ 
is given by
\begin{equation}
\mlqnearmaxsq = (m_\sql^2 - m_\ntltwo^2 )
                             (m_\ntltwo^2 - m_\sslr^2) / m_\ntltwo^2 \ .
\end{equation}

As one would expect, for events in which $\mlqnear$ is at its maximum value
$\sin^2(\theta^*/2) = 1$, \ie\ $\theta^*=\pi$, 
and the near lepton and the quark are back-to-back
in the rest frame of the \ntltwo.

If the spin correlations were ignored, and particles
were allowed to decay according to phase-space, then 
the probability density function would be
\begin{equation}
\frac{dP_\mathrm{PS}}{d\hat{m}} = 2\sin(\theta^*/2) = 2\hat{m}
\label{eq:Pps}
\end{equation}
where the rescaled invariant mass variable, $\hat{m}$, is defined by
\begin{equation}
\hat{m}\equiv \mlqnear/\mlqnearmax = \sin(\theta^*/2)\ .
\label{eq:mhat}
\end{equation}
Taking spin correlations into account, there are extra spin 
projection factors in the amplitude 
of $\sin(\theta^*/2)$ or $\cos(\theta^*/2)$ depending 
on the helicities of the \anti lepton and \anti quark, 
so that for 
$l^+ q$ or $l^-\bar{q}$
\begin{equation}
\frac{dP_1}{d\hat{m}} = 4 \sin^3(\theta^*/2)
                           = 4 \hat{m}^3
\label{eq:P1}
\end{equation}
while for $l^- q$ or $l^+\bar{q}$ 
\begin{equation}
\frac{dP_2}{d\hat{m}} = 4 \sin(\theta^*/2)\cos^2(\theta^*/2) 
                           = 4 \hat{m}(1-\hat{m}^2) \ .
\label{eq:P2}
\end{equation}
The functional form of these distributions is shown in \figref{fig:shapes}.

\section{Parton level}

%% \TABULAR[b]{|l|l|l|l|l|}{
%% \hline
%% \mnought & \mhalf & $A_0$ & $\tan(\beta)$ & $\mathrm{sign}(\mu)$\\
%% \hline
%% 100~GeV    & 300~GeV    & 300~GeV   & 2.1           & + \\
%% \hline
%% }{
%% \label{tab:point}
%% Definition of the GUT-scale parameters for the point investigated.
%% The top quark mass (important in the RGE running) was set to 175~GeV.
%% }

\begin{table}[b]\begin{center}\begin{tabular}{|l|l|l|l|l|l|l|l|l|l|}
\hline
\gluino & \ntlone & \ntltwo & \ssul & \ssdl & \sser & \ssel\\
\hline
717  & 116   & 213   &  631   & 634    & 153    & 229\\
\hline
\end{tabular}\caption{
\label{tab:mass}
Masses of selected particles (in GeV) for the model point investigated.
}
\end{center}
\end{table}

%% \EPSFIGURE{sugra_pt5.1200.in.eps, height=10cm, angle=+90}{
%% Part of the decay spectrum for the model point invesigated.
%% Solid black lines indicate branching ratios (BRs) greater than 10\%,
%% dashed blue lines show BRs in the range $1\% \to 10\%$, while
%% red dotted lines show BRs in the range $0.1 \to 1\%$.
%% The vertical axis shows masses in GeV.
%% The sparticles are displaced horizontally for clarity.
%% \label{fig:spectrum}
%% }

I take as an example model the mSUGRA point
with $m_0= 100$~GeV, $m_\half=300$~GeV, $A_0=$~300~GeV, 
$\tan\beta=2.1$ and $\mu>0$.
For this point the decay chain (\figref{fig:feyn}) has already been well
investigated in the context of sparticle mass 
measurements\cite{phystdr,Lester:2001zx}.
The point was originally motivated by cosmological 
\ntlone\ relic density considerations,
but it has subsequently been excluded 
by the LEP light Higgs limit\cite{Barate:2003sz}.
This has no bearing on the analysis presented in this \letter, 
since the point is used only to illustrate the method.

The SUSY mass spectrum and decay branching ratios were calculated with \isajetv{7.64}. 
Some of the important sparticle masses are shown in \tabref{tab:mass}.
These values differ from those of \cite{phystdr,Lester:2001zx}
because a more recent version of the \isajet\ program has been used
with updated renormalization group evolution.
One significant point to mention in respect to this
analysis is that the \ntltwo\ is largely Wino, 
so the branching ratios $\sqr\to\ntltwo q$ are
highly suppressed compared to the equivalent decays for the left-handed squarks.

\label{sec:lqnear}
To examine the parton-level distributions, a small sample of inclusive SUSY events 
$(\approx2.5~\invfb)$ was generated using 
the \herwigv{6.505} Monte Carlo event generator, with the 
leading-order parton distribution functions 
of {\tt MRST}\cite{Martin:1998sq} (average of central and higher gluon).
There is an obvious difference between the \lqnear\ distributions for leptons 
and anti-leptons (\figref{fig:near}a).
The differences between these shapes and those in the idealized 
distributions (\figref{fig:shapes}) are largely due to 
sparticles being off their mass shells and
contributions from the various different-mass squarks.  

The distinctive charge asymmetry is caused by spin 
correlations carried by the \mbox{spin-\half} \ntltwo, 
so if one could measure the \mlqnear\ invariant mass distribution
in the decay chain (\ref{eq:partchain}) 
it would be easy to show that the \ntltwo\ is \mbox{spin-\half}.

\EPSFIGURE{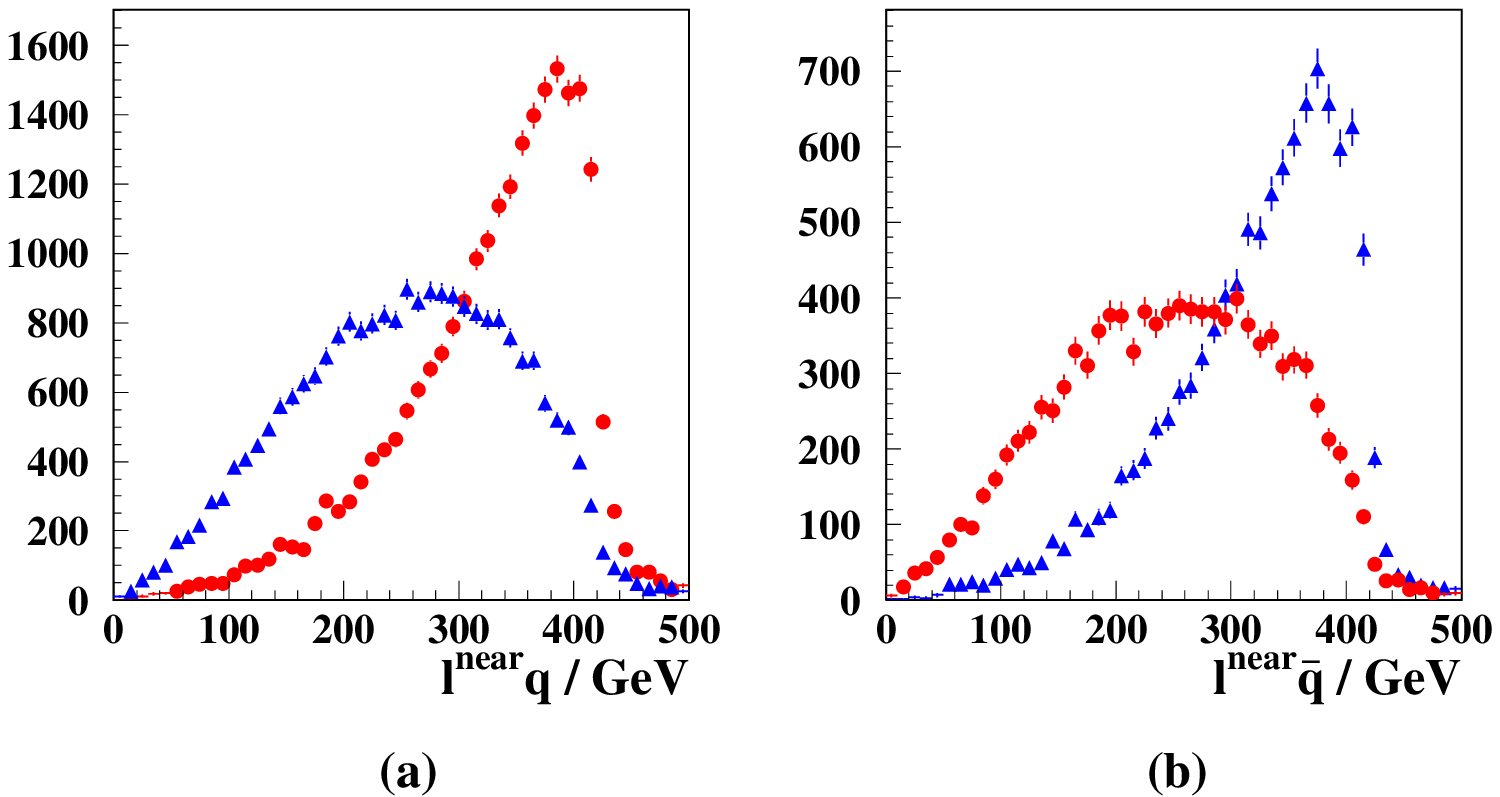, height=8.5cm}{
Invariant mass distributions of {\bf (a)} \lqnear\ and {\bf (b)} \lqbnear,
at the parton level.
The triangles are for a negatively charged near lepton, 
while the circles are for a positively charged near lepton.
For the test point the on-shell kinematic maximum is 413.4~GeV.
Note that these distributions cannot be measured 
directly by the experiment.
%$\int L dt\approx 2.5~\invfb$.
\label{fig:near}
}

However there are experimental difficulties in making such a measurement.
As was noted in \cite{Richardson:2001df}, in the decay of 
an anti-squark the asymmetry in the lepton charge distributions 
is in the opposite sense to that from squark decays (see \figref{fig:near}b).
As it is experimentally unlikely that one could distinguish between 
a quark and an anti-quark jet at the LHC, 
only the sum of the $lq$ and $l\bar{q}$ distributions 
can be considered to be observable.
Thus, if equal numbers of squarks and anti-squarks were produced,
the \lqnear\ distribution would be indistinguishable from the 
phase-space distribution, and no spin information could be obtained.
However the fact that the LHC will be a proton-proton collider 
means that the production of squarks will be enhanced compared to anti-squarks,
which can then lead to a significant spin-generated lepton charge asymmetry.

\section{$\tilde{q}-\tilde{\bar{q}}$ production asymmetry}

In a $pp$ collider, the production processes
\begin{equation}
q g \to \tilde{q} \tilde{g} \quad \mathrm{and} \quad \bar{q} g \to \tilde{\bar{q}} \tilde{g}
\label{eq:pro}
\end{equation}
will produce more squarks than anti-squarks. This is because the
quark parton distribution function (PDF) is larger than 
that of the anti-quark due to the presence of the valence quarks.

For a significant $\tilde{q}-\tilde{\bar{q}}$ asymmetry, the processes 
(\ref{eq:pro}) must provide a significant
contribution to the \anti squark production.
That is to say that they must not be much smaller than 
competing charge-symmetric processes such as
\begin{equation}
q \bar{q} \to \tilde{q} \tilde{\bar{q}} \quad , \quad 
g g \to \tilde{q} \tilde{\bar{q}}
\end{equation}
or gluino pair production followed by decay to \anti squarks.
One must also check that the processes (\ref{eq:pro})
sample the parton distribution functions 
in a region of $x$ and $\mu_F^2$ where valence quark PDFs are significant.

It is safe to assume that by the time SUSY spin measurements are 
being made at the LHC, the PDFs will be well measured
from other measurements, for example from electroweak boson production.
One would expect that the production asymmetry, which is the result of the 
well measured valence quark distribution,
would be insensitive to any remaining uncertainty in the PDFs.
%However the study of any remaining PDF dependence is beyond the scope of this \letter.

\begin{figure}
\begin{center}
  \begin{minipage}[b]{.49\linewidth}
   \begin{center}
    \epsfig{file=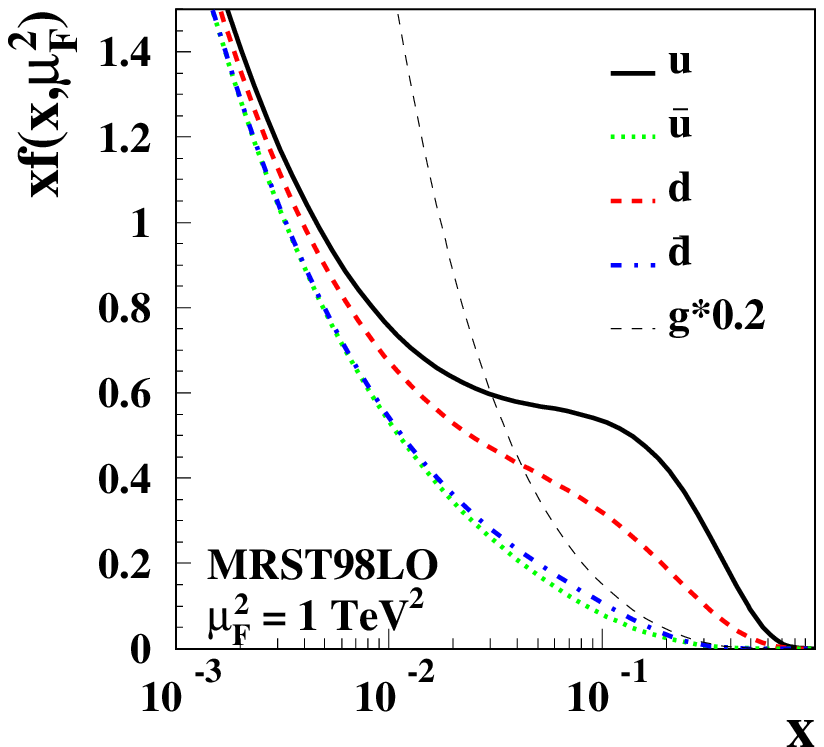, height=6cm}
     \\ \ \hspace{1cm} {\bf(a)}
    \end{center}
  \end{minipage}\hfill
  \begin{minipage}[b]{.49\linewidth}
    \begin{center}
     \epsfig{file=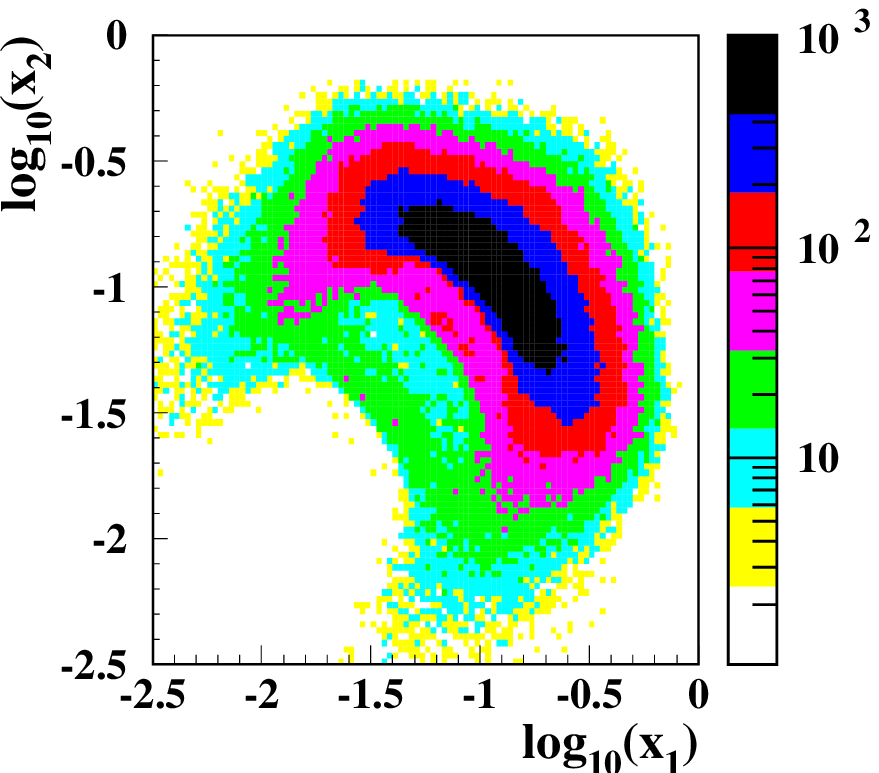, height=6cm}
     \\ \ \hspace{1cm} {\bf(b)}
    \end{center}
  \end{minipage}\hfill
\caption{
{\bf (a)} The parton distribution functions
used in this \letter\ plotted at factorization scale $\mu^2_F=1~\TeV^2$ \cite{Martin:1998sq}.
{\bf (b)} 2-dimensional histogram of the values of $x_1$ and $x_2$ sampled 
by the Monte Carlo in sparticle pair production at the mSUGRA point investigated.
\label{fig:pdf}
}
\end{center}
\end{figure}

The Bjorken $x_{1,2}$ parameters, giving the fraction of the proton momenta 
carried by the initial state partons satisfy
\begin{equation}
x_1 x_2 = M^2/s \ ,
\end{equation}
where $s$ is the usual Mandelstam variable.
For the LHC, $\sqrt{s}=14$~TeV, 
and for the test point the invariant mass, $M$, 
for squark pair production is about 1.3~TeV,
so the $x_i$ can be expected to be of the order of 0.1. This is confirmed
from the values of $x$ sampled by the Monte Carlo (\figref{fig:pdf}b).

In terms of producing a $\tilde{q}-\tilde{\bar{q}}$ asymmetry, 
it is advantageous that at $x\approx 0.1$ and $\mu_F^2\sim$~1~TeV$^2$ 
the valence quark parton distribution functions are large (\figref{fig:pdf}a).
The over-production of squarks relative to anti-squarks 
for the test point can be observed by comparing the 
normalisation of the $lq$ and $l\bar{q}$ plots in \figref{fig:near}, where 
it can be seen that approximately twice as many squarks are produced 
as anti-squarks for this point.

\section{Experimental observables}
\EPSFIGURE{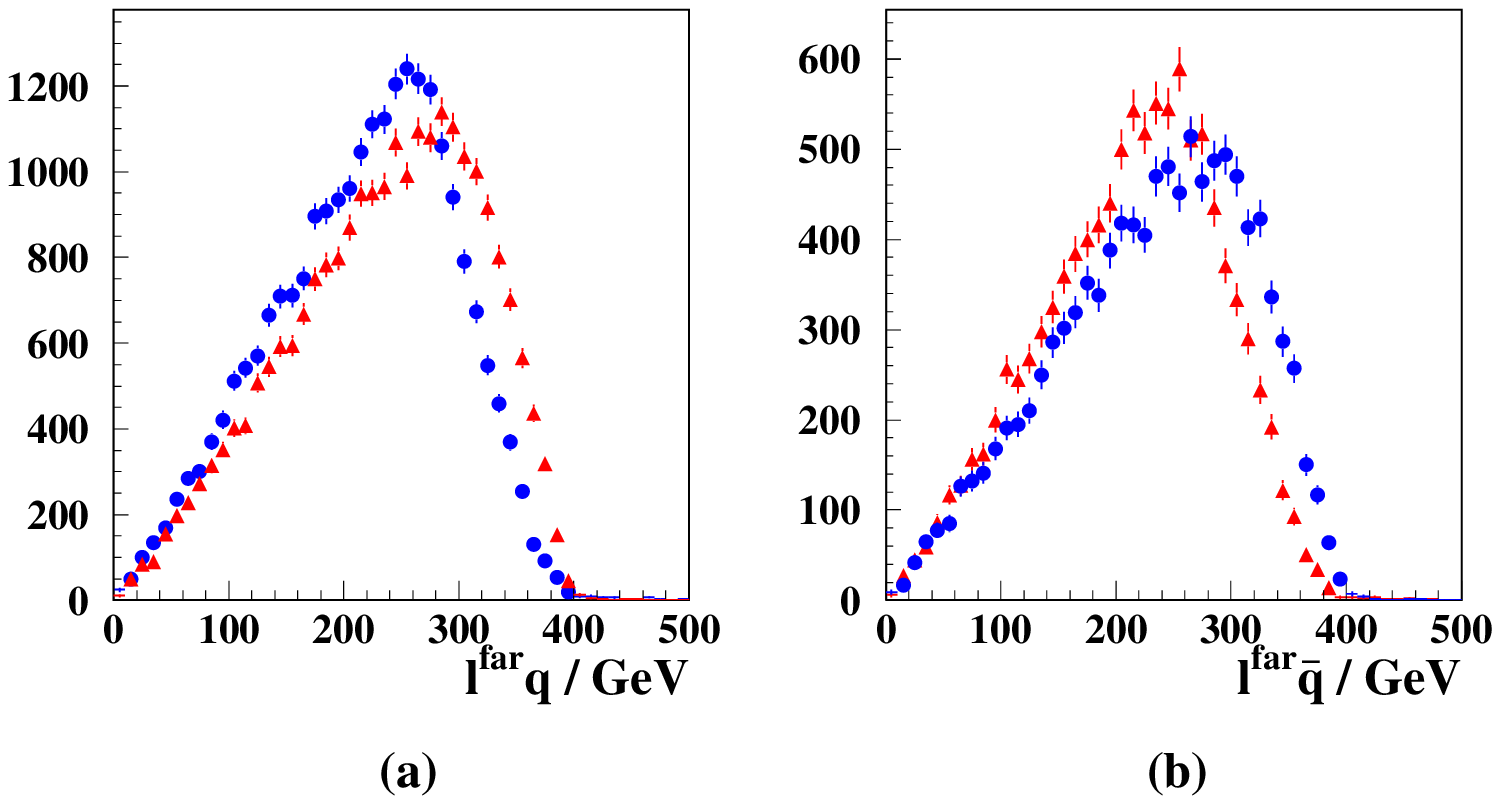, height=8.5cm}{
Invariant mass distributions of {\bf (a)} \lqfar\ and {\bf (b)} \lqbfar,
at the parton level.
The circles indicate the distribution for the negatively
charged far lepton, while the triangles are for 
the positively charged far lepton.
Note that these distributions, like \figref{fig:near}, 
cannot be directly measured by the experiment.
The explanation for the charge asymmetry is given in the text.
%$\int L dt\approx 2.5~\invfb$.
\label{fig:far}
}

Although the \lqnear\ invariant mass would provide the theoretically cleanest spin signature,
with the experimental data it will not be possible to distinguish the 
near lepton (from the \ntltwo\ decay) from the
far lepton (from the \sslrpm\ decay) on an event-by-event basis.
Since all of these signal events will contain two leptons of the
same family but of different sign, what one {\em can} do experimentally
is to look for asymmetries in the \lqplus\ and \lqminus\ distributions,
each of which will contain contributions from both the near and the far 
lepton.

The asymmetry in the \lqnear\ invariant mass was discussed in \secref{sec:lqnear}.
One might naively assume that the invariant mass distribution of the 
far \anti lepton with the \anti quark would be free 
from spin-correlation effects, since this lepton originates 
from the decay of a scalar particle, 
the \sslrpm, which should wash out any spin effects.
However the slepton itself has been produced in the decay of the 
\ntltwo, and so has a boost relative to the quark jet which 
depends on its charge. 
To be explicit about the sense of the difference, 
recall that the near positive lepton favours being back-to-back with 
a $q$ jet. This decreases the boost of the negative slepton
relative to the $q$ jet, and means that on average the invariant mass of the 
quark with a negative far lepton will be smaller than with a positive lepton.
The \lqfar\ invariant mass distributions 
are shown in \figref{fig:far}a while the equivalent distributions 
from anti-squark decay, \mlqbfar, are plotted in \figref{fig:far}b.

\begin{figure}
\begin{center}
  \begin{minipage}[b]{.49\linewidth}
   \begin{center}
    \epsfig{file=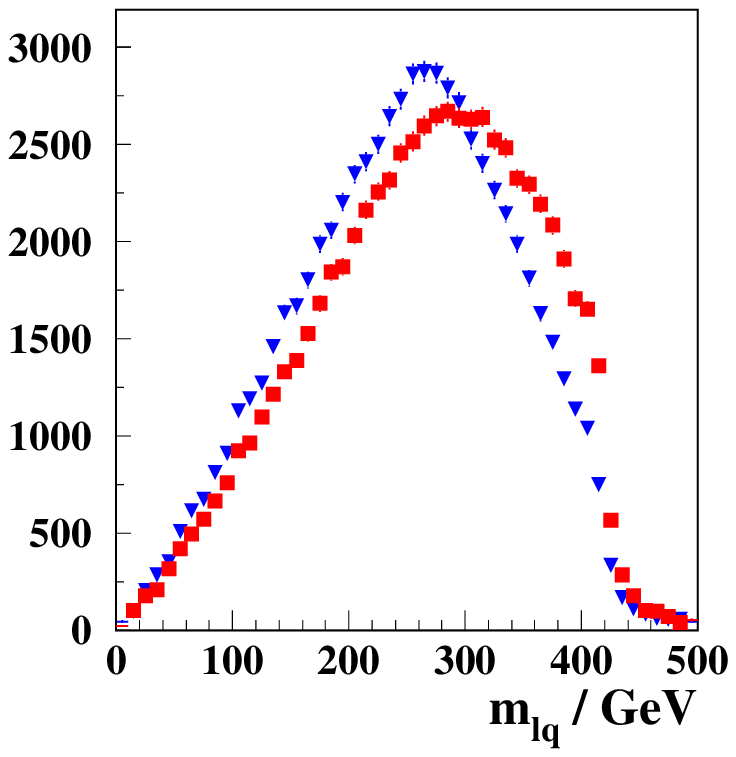, width=6.6cm}
     \\ \ \hspace{1cm} {\bf(a)}
    \end{center}
  \end{minipage}\hfill
  \begin{minipage}[b]{.49\linewidth}
    \begin{center}
     \epsfig{file=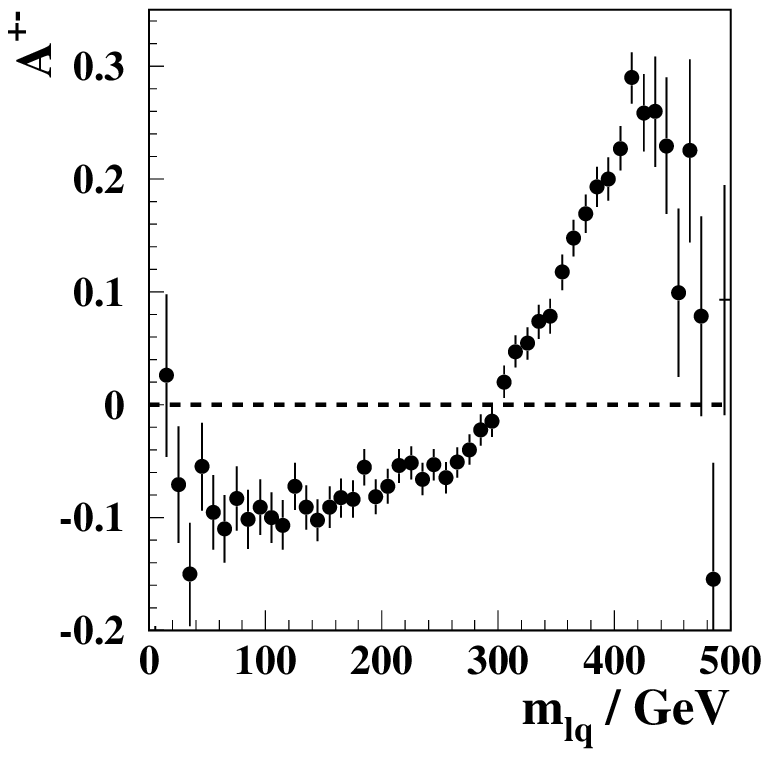, width=6.6cm}
     \\ \ \hspace{1cm} {\bf(b)}
    \end{center}
  \end{minipage}\hfill
\caption{
{\bf (a)} The $l^+ q$ (squares) and $l^- q$ (triangles) 
invariant mass distributions, and 
{\bf (b)} the charge asymmetry $A^{+-}$ (eq. \ref{eq:asym}) at the parton level.
These distributions have folded-in the indistinguishability 
of the near and far leptons, and quark vs. anti-quark jets.
%$\int L dt\approx XXX~\invfb$.
\label{fig:asym}
}
\end{center}
\end{figure}

The \lqplus\ and \lqminus\ distributions contain 
contributions from both near and far leptons,
and from squark and anti-squark decays.
The spin information carried by the \ntltwo\
causes an obvious difference in shape between
these experimentally-accessible distributions (\figref{fig:asym}a).
The charge asymmetry in the differential cross-sections
is defined here as
\begin{equation}
A^{+-} \equiv \frac{s^+ - s^-}{s^+ + s^-},
\quad \mathrm{where}\quad s^\pm = \frac{d\sigma}{d(m_{l^{\pm}q})}\ .
\label{eq:asym}
\end{equation}
\label{sec:plusminus}
This asymmetry, plotted as a function of $m_{lq}$ in \figref{fig:asym}b,
is clearly not consistent with zero.
This demonstrates that at parton level the 
asymmetry survives contamination from anti-squark
production, and the experimental lack of knowledge
of near vs. far lepton.

%\TABULAR[b]{|l|l|l|l|}{
%\hline
%\multicolumn{2}{|c}{unobservable} & 
%\multicolumn{2}{|c}{observable} \vline\\
%\hline
%\lqnear & \lqfar & \lqhigh & \lqlow\\
%\hline
%413.4 & 388.6 & 413.4 & 325.1\\
%\hline
%}{
%Maximum values (in GeV) for the various lepton-quark invariant mass distributions,
%using results from \cite{Lester:2001zx}.
%I have used the mass of the \ssul\ to represent the \sql\ masses.
%\label{tab:edges}
%}

\begin{figure}
\begin{center}
  \begin{minipage}[b]{.49\linewidth}
   \begin{center}
    \epsfig{file=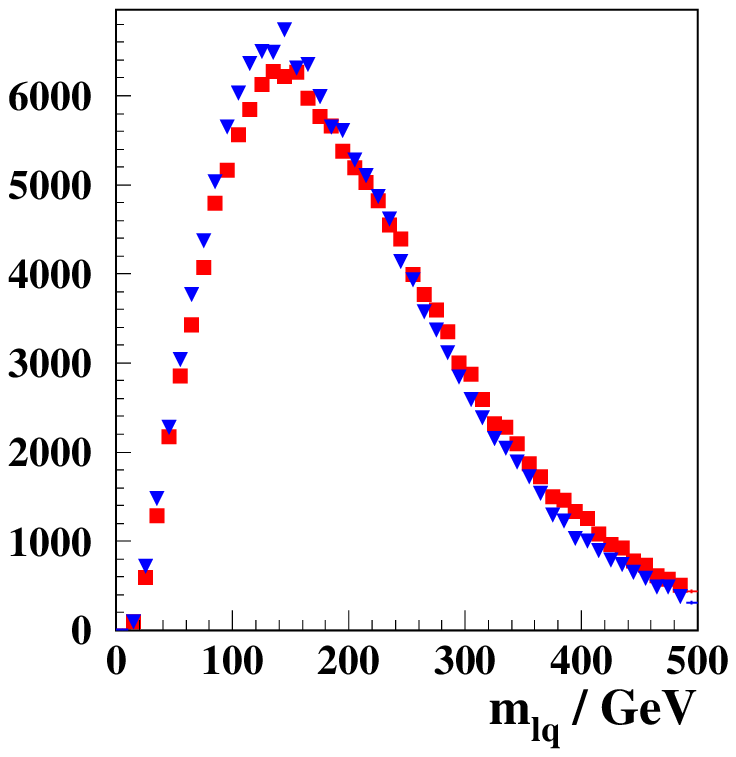, width=6.5cm}
     \\ \ \hspace{1cm} {\bf(a)}
     \\ \epsfig{file=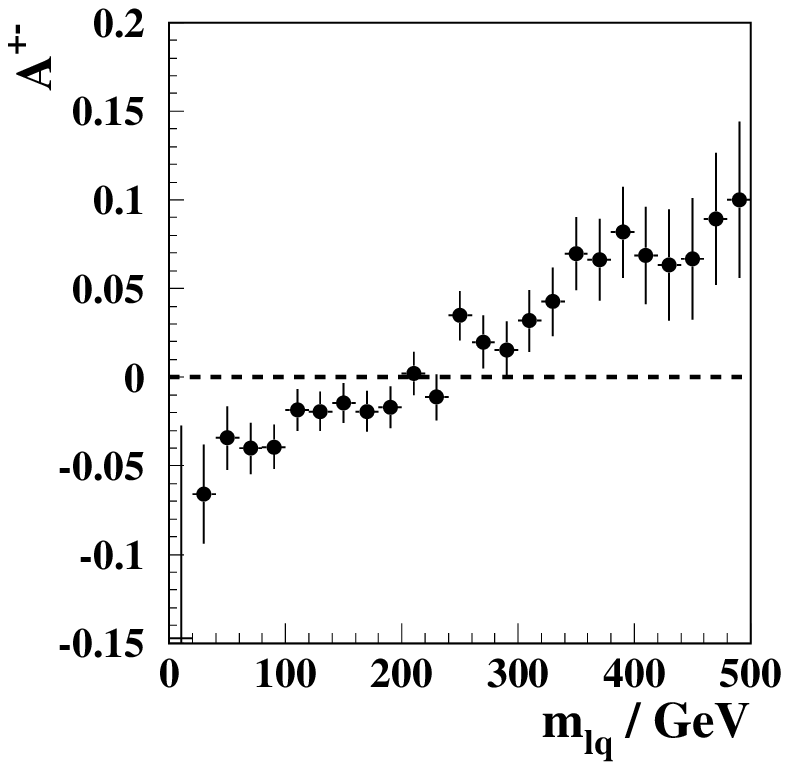, width=6.5cm}
     \\  \ \hspace{1cm} {\bf(c)}
    \end{center}
  \end{minipage}\hfill
  \begin{minipage}[b]{.49\linewidth}
    \begin{center}
     \epsfig{file=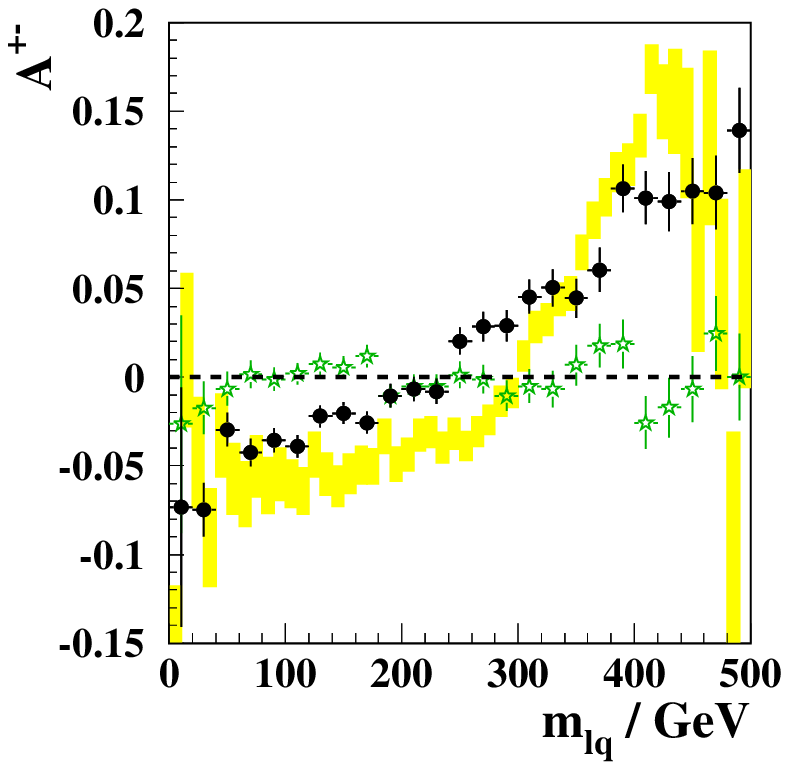, width=6.5cm}
    \\  \ \hspace{1cm} {\bf(b)}
    \\     \epsfig{file=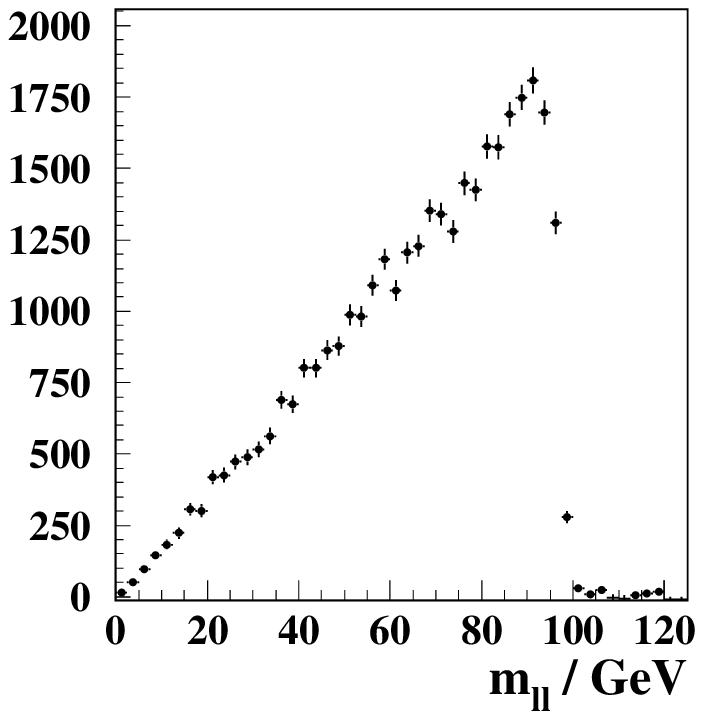, width=6.5cm}
    \\  \ \hspace{1cm} {\bf(d)}
    \end{center}
  \end{minipage}\hfill
\caption{
{\bf (a)} The \lqplus\ (squares) and 
\lqminus\ (triangles) invariant mass distributions
after detector simulation and event selection.
{\bf (b)} The solid circles show the lepton charge asymmetry $A^{+-}$ %(\eqref{eq:asym})
as a function of $m_{lq}$, again after detector simulation.
The shaded rectangles are the parton-level result scaled down by a factor of 0.6.
The stars show a cross-check -- the equivalent detector-level asymmetry 
with spin correlations suppressed. 
For both of the upper two plots $\int Ldt$~=~500~\invfb.
%{\bf (c)} The chi-squared divided by the number (25) of degrees of freedom
%as a function of integrated luminosity collected,
%for the detector-level asymmetry distribution. The red circles
%correspond to the simulation with full spin correlations,
%while the blue triangles are for the phase-space cross-check.
%The horizontal dotted lines show the confidence with which 
%the null hypothesis of no asymmetry could be excluded.
{\bf (c)} The detector-level charge asymmetry, $A^{+-}$, 
with spin correlations, using a 150~\invfb\ subset of the data.
{\bf (d)} The opposite-sign, same-family dilepton invariant mass 
distribution after opposite-sign, different-family subtraction.
\label{fig:atlasym}
}
\end{center}
\end{figure}

To check that this asymmetry is robust, a 500~\invfb\ sample of inclusive SUSY events 
was generated with \herwig\ at the same mSUGRA point.
The events were then passed through the \atlfastv{2.50}\cite{Richter:1998at} 
detector simulation.

The cuts applied are taken directly from a previous analysis\cite{Lester:2001zx}
in which it was emphasised that this selection had not been
tuned to the particular mSUGRA point under investigation,
and so could claim a degree of model-independence.

In brief, the event selection requires:
\begin{itemize}
\item{
exactly 2 electrons or muons of the same family with opposite charge, both having $p_T\geq10$~GeV.
}\item{
four or more jets with $p_T^{j_1} \geq 100$~GeV and 
$p_T^{j_k}\geq$~50~GeV for $k \in \{2,3,4\}$, where $p_T^{j_i}$ is the transverse momentum
of the $i$th jet ordered in $p_T$ such that $p_T^{j_1}>p_T^{j_2}>p_T^{j_3}>p_T^{j_4}$.
}\item{
missing transverse momentum, $\ptmiss \geq \max(100~\GeV, 0.2~M_\mathrm{effective})$ 
and $M_\mathrm{effective}\geq 400~\GeV$, where 
$M_\mathrm{effective} = \ptmiss + \Sigma_{i=1}^4 p_T^{j_i}$
}\item{
dilepton invariant mass, $m_{ll}\leq m_{ll}^{\max} + 1~\GeV$
}\item{
dilepton plus jet invariant mass, $m_{llq}\leq m_{llq}^{\max}$ 
where $m_{llq} = \min(m_{llj_1}, m_{llj_2})$.
}
\end{itemize}
A more detailed discussion of this selection, including the
definition of the kinematic limits $m_{ll}^{\max}$ 
and $m_{llq}^{\max}$ is offered in \cite{Lester:2001zx}.
After applying these cuts the remaining Standard Model background
has been shown to be much smaller than the SUSY background\cite{phystdr}, 
and so has not been simulated here.

For real data, one would expect that charge-blind 
$m_{l^\pm q}$ and other invariant mass distributions would be investigated  first. 
One would use these to extract information on the mass of the various sparticles.
Using this information, the selection parameters could be tuned
before making any attempt to measure the lepton charge asymmetry,
however no such tuning has been attempted for this \letter.

While the shapes of the $m_{l^\pm q}$ distributions (\figref{fig:atlasym}a) 
are modified by detector simulation and event selection, 
the cuts are charge-blind, and so the 
detector-level asymmetry function (\figref{fig:atlasym}b)
retains the same approximate shape as at parton level.
To help visual comparison the parton-level asymmetry has been 
scaled down by a factor of 0.6. A decrease in the observed asymmetry
can be expected due to pollution from the SUSY background. 
A more sophisticated analysis could
estimate this background as a function of $m_{lq}$ using a
same-sign dilepton event selection.
Some smearing is apparent, but the fact that the asymmetry
is negative at low \mlqnear\ and positive at high \mlqnear\ is very clear.
The shape of the asymmetry function therefore strongly favours a spin-\half\ \ntltwo.

A nice contrast can be seen by making a cross-check in which
the spin correlations have been switched off in the Monte Carlo
so that all particles decay according to pure phase-space.
This produces no significant asymmetry (\figref{fig:atlasym}b)
and reflects in some sense the experimental expectation for a `scalar \ntltwo'.

\begin{figure}
\begin{center}
\caption{
}
\end{center}
\end{figure}

Figure \ref{fig:atlasym}b was made with a high-statistics 
(500~\invfb) sample so that the shape of the distribution could be clearly seen.
However such a large sample is not necessarily required to determine that the
spin correlations exist.
%In \figref{fig:atlasym}c, the goodness of fit to the null hypothesis of no
%spin correlations is shown as a function of integrated luminosity.
The integrated luminosity required to make the measurement will be different for 
each SUSY point, depending on the sparticle spectrum.
However for the particular mSUGRA point investigated, 
the spin-generated charge asymmetry can be seen to be inconsistent with zero
with 150~\invfb\ of integrated luminosity (\figref{fig:atlasym}c).

It is worth adding that further spin information is available 
from the dilepton invariant mass distribution. 
This also has a kinematic limit, this time at 
\begin{equation}
(m_{l^+l^-})^2_\mathrm{max} = (m_\ntltwo^2 - m_\sslr^2 ) 
                               (m_\sslr^2 - m_\ntlone^2) / m_\sslr^2 \ .
\end{equation}
Since one lepton of either sign is produced, 
and both are right-handed, one might expect that the 
angular distribution would show strong spin effects.
However the scalar slepton removes all
spin correlations between the leptons,
and so the $m_{l^+l^-}$ distribution (\figref{fig:atlasym}d) 
is in very good agreement with the triangular prediction of phase-space (\figref{fig:shapes}). 
This agreement would be hard to explain except 
as a result of a heavy scalar particle carrying lepton number,
and so would increase confidence in the supersymmetric nature
of the particles participating in the decay chain.

\section{Conclusions}

The method presented shows that it is possible to determine
the spins of some of the SUSY partners of Standard Model 
particles at the LHC.
The charge asymmetry lends good supporting evidence
to the hypothesis that one is observing supersymmetry 
and to the identity of the sparticles participating in the decay chain.
For the point examined, the shape of the asymmetry 
distribution could be observed with an integrated luminosity of 150~\invfb.

It should be noted that the method presented here requires
a particular decay chain (\figref{fig:feyn}) to occur at a reasonably
high rate, and an initial asymmetry in the 
squark vs. anti-squark production cross-sections.
However, it could be employed to disentangle supersymmetry from 
other phenomenologically interesting models,
such as universal extra dimensions with 
Kaluza-Klein parity\cite{Cheng:2002ab}.

%stau measurements?
%may help in distinguishing near and far edges?

\section*{Acknowledgments}
I would like to thank my colleagues in the Cambridge SUSY working group
for helpful discussions, particularly Bryan Webber and Chris Lester.
I am also pleased to thank Robert Thorne, and the members of the Oxford ATLAS particle 
physics group, especially Claire Gwenlan,
for discussions relating to parton distribution functions.
I have made use of the physics analysis framework and tools which are
the result of ATLAS collaboration-wide efforts.
This work was funded by PPARC.

\bibliography{thesis}

\begin{thebibliography}{1}
\expandafter\ifx\csname url\endcsname\relax
  \def\url#1{\texttt{#1}}\fi
\expandafter\ifx\csname urlprefix\endcsname\relax\def\urlprefix{URL }\fi

\bibitem{Richardson:2001df}
P.~Richardson, JHEP 11 (2001) 029.

\bibitem{Corcella:2000bw}
G.~Corcella, et~al.,
JHEP  01 (2001) 010.

\bibitem{Moretti:2002eu}
S.~Moretti, K.~Odagiri, P.~Richardson, M.~H. Seymour, B.~R. Webber,
  JHEP 04 (2002) 028.

\bibitem{Lester:2001zx}
C.~G. Lester,
Ph.D.  thesis, {CERN}-{THESIS}-2004-003 (2002).

\bibitem{phystdr}
{ATLAS} Detector and Physics Peformance TDR, CERN, 1999.

\bibitem{Barate:2003sz}
R.~Barate, et~al.,
Phys. Lett. B565 (2003) 61--75.

\bibitem{Martin:1998sq}
A.~D. Martin, R.~G. Roberts, W.~J. Stirling, R.~S. Thorne,
Eur. Phys. J. C4 (1998) 463--496.

\bibitem{Richter:1998at}
E.~Richter-Was, D.~Froidevaux, L.~Poggioli,
 {ATLAS} note {ATL}-{PHYS}-98-131.

\bibitem{Cheng:2002ab}
H.-C. Cheng, K.~T. Matchev, M.~Schmaltz, 
Phys. Rev. D66 (2002) 056006.

\end{thebibliography}

\end{document}